\documentclass[%
mathleft,%
 earlyview,%
]{an}

\usepackage{natbib}
\usepackage{graphicx}
\usepackage{times}

\usepackage{journals}
\bibpunct{(}{)}{;}{a}{}{,}

\newcommand{\ms}{m\,s$^{-1}$}

\begin{document}

\Pagespan{1}{}
\Yearpublication{2014}%
\Yearsubmission{2013}%
\Month{1}%
\Volume{335}%
\Issue{1}%
\DOI{This.is/not.aDOI}%

\title{HERMES at Mercator, competitive high-resolution spectroscopy with a small telescope
    \thanks{Based on observations made with the Mercator Telescope,
    operated on the island of La Palma by the Flemish Community, at the
    Spanish Observatorio del Roque de los Muchachos of the Instituto de
    Astrof\'{i}sica de Canarias.}
}

\author{Gert Raskin
  \thanks{Corresponding author
  \email{Gert.Raskin@ster.kuleuven.be}}
\and  Hans Van Winckel
}
\titlerunning{HERMES at Mercator, competitive high-resolution spectroscopy with a small telescope}
\authorrunning{G. Raskin}
\institute{
Instituut voor Sterrenkunde, KU\,Leuven, Celestijnenlaan 200D, B-3001 Leuven, Belgium}

\received{XXXX}
\accepted{XXXX}
\publonline{XXXX}


\abstract{HERMES, a fibre-fed high-resolution (R\,=\,85000) echelle spectrograph with good stability and excellent throughput, is the work-horse instrument of the 1.2-m Mercator telescope on La Palma. HERMES targets building up time series of high-quality data of variable stellar phenomena, mainly for asteroseismology and binary-evolution research.
In this paper we present the HERMES project and discuss  the instrument design, performance, and a future upgrade. We also present some results of the first four years of HERMES observations. We illustrate the value of small telescopes, equipped with efficient instrumentation, for high-resolution spectroscopy.  
}

\maketitle

\section{Introduction}
The Mercator telescope (\texttt{www.mercator.iac.es}, \citealt{raskin04}) is a semi-automatic \mbox{1.2-m} telescope, installed at the Roque de los Muchachos Observatory on La Palma (Canary Islands, Spain). It is funded by the Flemish Community of Belgium and operated by the Institute of Astronomy of the University of Leuven (KU\,Leuven, Belgium). We built this telescope, a copy of the Swiss Euler telescope in La Silla (Chile), in collaboration with the Observatoire de Gen\`{e}ve (Switzerland). The main asset of Mercator is its permanent long-term availability. This gives  the telescope a specific niche in today's era of ever growing telescope dimensions.  Indeed, it allows our astronomers to embark on intensive monitoring campaigns covering a broad range of time scales, extending from minutes to years. Furthermore, flexible time allocation and scheduling  allow us to quickly seize emerging opportunities and obtain immediate follow-up observations to complement interesting data from other facilities. For obvious reasons, both flexibility and long-term availability are difficult to attain from large telescopes. Given the size of the telescope, the main research theme is in the domain of stellar astrophysics.

This avail of small telescopes is even more relevant for high-resolution spectroscopy. In this field, large telescopes require small sky apertures or very large instruments to obtain high spectral resolution. The larger sky apertures of small telescopes allow them to pursue observing in an  efficient way, even under very moderate seeing conditions. Moreover, instrument design for small telescopes usually suffers less from the broad variety of constraints that tends to accompany large facility projects.  Efficiency and throughput can only benefit from reduced design constraints. 

On the other hand, with a small telescope it is even more important to pay utmost attention to instrumental efficiency in order to be able to do relevant and competitive science. This is what we have tried to achieve with HERMES (an acronym for High Efficiency and high Resolution Mercator Echelle Spectrograph): design and build a high-resolution spectrograph with very high efficiency. At the same time, we also required HERMES to have excellent stability in order to obtain accurate radial velocity measurements. Good efficiency is not only limited to high throughput but it also includes broad wavelength coverage in a single exposure (380\,--\,900\,nm) as well as efficient exploitation of the telescope and spectrograph combination. The latter is particularly important for monitoring programs and time-critical observations. That is also the reason why most of the observations with  Mercator are  scheduled from a common pool. Especially for this purpose, we developed software that assists the observer in scheduling an optimised observing night  \citep{merges12}, based on observing program priorities, timing constraints and weather conditions. 

An international consortium was set up to for the realisation of HERMES. The Institute of Astronomy of the KU\,Leuven (Belgium) headed  the project and carried out the major part of the work. Consortium partners were the Institut d'Astronomie et Astrophysique (ULB,
Belgium) and the Royal Observatory of Belgium, both contributing with the data-reduction pipeline, the Th\"uringer Landessternwarte Tautenburg (Germany) building an important part of the spectrograph mechanics, and the Geneva Observatory (Switzerland) in charge of the telescope interface mechanics. The HERMES project already kicked off in 2005. Immediately after  commissioning at the telescope in 2009, HERMES became the work-horse instrument of the Mercator telescope, being used during 80\% of the observing nights.

A description of HERMES was already given in \citet{Raskin11}. In this contribution, we present additional information about the design process and give extra details about some of the spectrograph key components and design choices. Further, we discuss a future upgrade of the fibre link and give an update of the instrument performance, including modal fibre noise measurements and signal-to-noise ratio limitations. Finally, we also present a few interesting science results that were obtained from HERMES spectra. 

\section{Spectrograph design}

\subsection{Layout}
Most nowadays high-resolution spectrographs are based on a \textit{white-pupil} (WP) layout \citep{baranne88}.
However, many excellent instruments with a different layout prove that a white pupil is not the only road to good performances for a high-resolution spectrograph. For that reason,  we performed a trade-off study to compare the performance of a spectrograph based on a WP and one using a folded  Schmidt (FS)  camera that, at the same time, serves as collimator, like e.g. SOPHIE \citep{Perruchot08}. Both spectrographs rely on prisms for  cross dispersion. The WP uses two prisms while the FS needs only one in double pass. The outcome of this analysis is summarised below.

\begin{itemize}
 \item \textbf{Efficiency} The number of reflecting surfaces is about the same for both designs, however, central obscuration ($\sim$25\%) gives a clear disadvantage to the FS. This is only partially offset by the throughput losses of the additional lenses in the camera of a WP ($\sim$10\%).

 \item \textbf{Resolution} For a given maximum grating width, resolution is proportional to  $\sin\delta$ ($\delta$\,=\,blaze angle). This gives the WP a small resolution gain of 9\% (R4) or 6\% (R3) over the regular R2 grating that obligatory has to be used with the FS.

 \item \textbf{Slit rotation} A very small off-plane angle (0.8$^{\circ}$) is sufficient to separate the incident and dispersed beams of a WP. This keeps the total slit rotation small, despite a large $\delta$. The variable slit rotation of the FS, caused by the pre-dispersion of the prism in front of the echelle grating,  poses an important inconvenience for the data-reduction. This is especially important when an image slicer is used.

 \item \textbf{Cross dispersion} Both the larger dispersion (due to larger $\delta$) and the single-pass require stronger cross dispersion or larger prisms for the WP. However, both cost and efficiency wise, this is largely offset by the much larger beam diameter of the FS.

 \item \textbf{Stray light} emerging from the grating is an important issue in an echelle spectrograph. Shielding the detector from this stray light is complicated in a FS but can be done very efficiently in a WP spectrograph. Hence, the WP delivers cleaner spectra than the FS. 

 \item \textbf{Stability} In the FS design, all elements with optical power can be made of low thermal expansion material like Zerodur, increasing the thermal stability. The refractive camera of the WP design will inevitably suffer from some thermal sensitivity.

 \item \textbf{Cost} The FS requires larger and thus more expensive optical elements than the WP. On the other hand, the WP requires a more expensive refractive camera. After consultation with potential manufacturers, the FS appeared to be marginally more expensive than the WP.

 \item \textbf{Risk} Both the large corrector plate and the very large cross-disperser prism of the FS pose serious procurement risks. Few manufacturers showed interest in \mbox{manufacturing} the large  aspheric corrector plate.
\end{itemize}

\begin{figure*}
\includegraphics[angle=0, width=\linewidth]{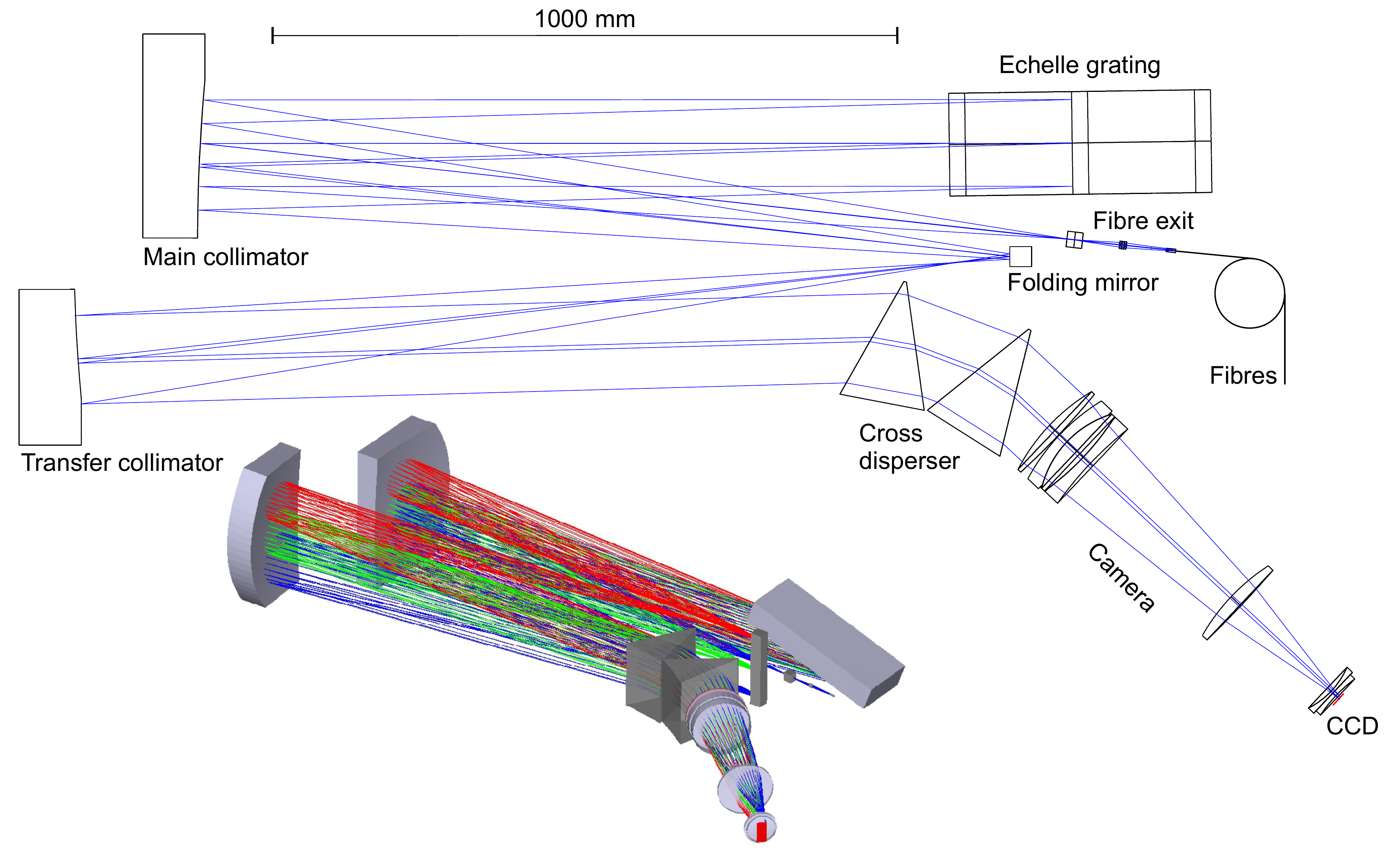}
\caption{\label{fig:layout} Ray-tracing and 3D-view of the white-pupil layout of the HERMES spectrograph.}
\end{figure*}

With the exception of the stability argument, the WP clearly outperforms the FS spectrograph. Hence, the design of HERMES (Fig.\,\ref{fig:layout})  is also  based on a WP layout.
 
\subsection{Dispersion}
Large wavelength coverage in a single exposure dictates the use of an echelle grating as main dispersion element. High efficiency over a large wavelength range can only be obtained by using prisms as cross disperser.
The echelle grating of first choice was a steep R4 echelle (blaze angle $\delta$\,=\,76$^{\circ}$) because it allowed for the smallest beam diameter for a given spectral resolution. However, we finally selected an R2.7 echelle ($\delta$\,=\,69.7$^{\circ}$) because it is offered with a finer groove spacing (52.7\,mm$^{-1}$) than is  available for the R4. The short spectral extent of the orders from coarser groove spacings have multiple disadvantages: 
\begin{itemize}
\item
Shorter orders require stronger cross dispersion and thus larger prisms in order to obtain  sufficient inter-order separation. On top of that, the stronger dispersion of the R4 required an additional proportional increase of the cross dispersion, leading to impractically large prisms.
\item
A 4k\,x\,4k detector would have been required to cover with sufficient sampling the full wavelength range in many short spectral orders. During the design phase of HERMES (2005--2006), this type of detectors was not yet very common and still too expensive for our budget. The long orders of the  52.7\,mm$^{-1}$ echelle nicely fit a more compact 2k\,x\,4.5k CCD.
\item
Many short orders complicate the data-reduction software and cause more spectral discontinuities in case of non-perfect order merging.
\end{itemize}

The HERMES echelle is a replica of the MR138 master grating form Richardson Gratings (Newport, USA) with a ruled area of 154\,x\,408\,mm. It covers the spectral range from 377 to 901\,nm in 55 orders.
Two identical 37.4$^{\circ}$ prisms provide sufficient cross dispersion to separate all overlapping echelle orders. The prisms consist of PBL1Y glass, a very light flint with high UV transmission  from OHARA's I-Line catalogue. They were manufactured and 1.2\% anti-reflection coated by Optique Fichou (France). The minimum order separation is 24 pixels at 900\,nm, increasing to 55 pixels at 380\,nm.

\subsection{Collimators and camera}
Both HERMES collimators were cut from a single 660-mm diameter Zerodur parabolic mirror. It has a focal length of 1400\,mm and the 152-mm diameter spectrograph beam leads to an aperture of \textit{f}/9.2. A blue-enhanced protected silver coating (Spectrum\,Thin\,Films, USA) gives good reflectivity  over most of the spectral range (Refl.\,$>$\,96\% at 400\,nm and  Refl.\,$>$\,99\% over 450 to 650\,nm) but sub-specification performance in the near-infrared (Refl.\,$>$\,94\% @ 800\,nm).

A refractive camera focusses the spectrum on the detector. Despite the moderate complexity of the camera (three singlets and one large triplet), it delivers excellent  image quality with 80\% encircled energy diameters substantially less than one detector pixel over the entire field of view and spectral range.  This could be achieved by sacrificing the correction of transverse chromatic aberrations. As a consequence, camera testing became  more complicated  and the camera will only be suitable for use in a spectrograph.  The last surface of the field flattener lens has a cylindrical shape to correct the dispersive field curvature. This field lens also acts as vacuum interface of the detector cryostat to avoid an additional plane window. The camera focal length of 475\,mm (\textit{f}/3.1)  gives a sampling of 2.3\,pixels of a high-resolution bin at the centre of the spectral orders. Up to 838\,nm, all orders are recorded  without gaps. The three reddest orders have small gaps at their extremes. The camera as well as the collimators were manufactured by SESO (France).

\subsection{Detector}
The HERMES detector is a standard thinned back-illuminated CCD with 2048\,x\,4608 13.5-$\mu$m pixels from e2v (UK). The spectral format of HERMES is very well matched to the detector geometry with an excellent ratio of wavelength coverage over detector area. HERMES was the first HR spectrograph to use a graded anti-reflection coating on the detector \citep{Kelt06}. These graded AR coatings have a thickness profile that is matched to the fixed spectral format of the spectrograph. This makes that the spectral characteristics of the coating are locally optimised for the wavelengths incident on each part of the detector. The thickness profile follows the cross-order gradient and the curved pattern of the spectral orders. 

Intuitively, one expects this type of coating to increase the quantum efficiency (QE) of the CCD. However, our experience could not fully confirm this. In the red part of the spectrum, QE is indeed higher than that of a typical CCD with a standard broad-band coating. In the green, we do not see a noticeable difference and for blue wavelengths, the performance of the graded coating, although compliant with the minimum specifications, is slightly smaller. Newer CCDs with the same type of graded coating were reported to perform better at blue wavelengths.

A second effect of the graded coating is that the reduced surface reflections also decrease fringing at NIR wavelengths. Indeed we measured a reduction of the fringing amplitude at 900\,nm by a factor of almost 10, compared to a detector with a standard broad-band coating.

\subsection{Fibre link}
Optical fibres link the telescope to the spectrograph. Two fibres in the focal plane of the telescope correspond with two observing modes for HERMES:
\begin{itemize}
\item
\textbf{HRF} (high-resolution fibre) offers high spectral resolution (R\,$\sim$\,85\,000) and is at the same time optimised for the high efficiency. This is achieved through a comfortably large sky aperture (2.5\,arcsec) and an image slicer.  To improve throughput, some focal ratio degradation in the fibre is accepted ($\sim$25\%) without overfilling the grating and loosing light in the spectrograph
\item
\textbf{LRF} (low-resolution fibre) still has a resolution of 63\,000 with a sky aperture of  2.15\,arcsec. It is optimised for high stability and accurate radial velocity measurements. Therefore, it can be used with the simultaneous Thorium technique, interlacing the LRF spectrum with that from a wavelength reference source, to track the instrumental drift during the exposure. An optical scrambler exchanges far and near fields of the two extremes of LRF, giving almost complete angular and radial scrambling, this to improve the illumination stability of the spectrograph. Better scrambling comes at the cost of smaller throughput. Losses in the scrambler result in a 30\% lower efficiency for LRF compared to HRF.
\end{itemize}

\begin{figure}
\includegraphics[angle=0, width=\linewidth]{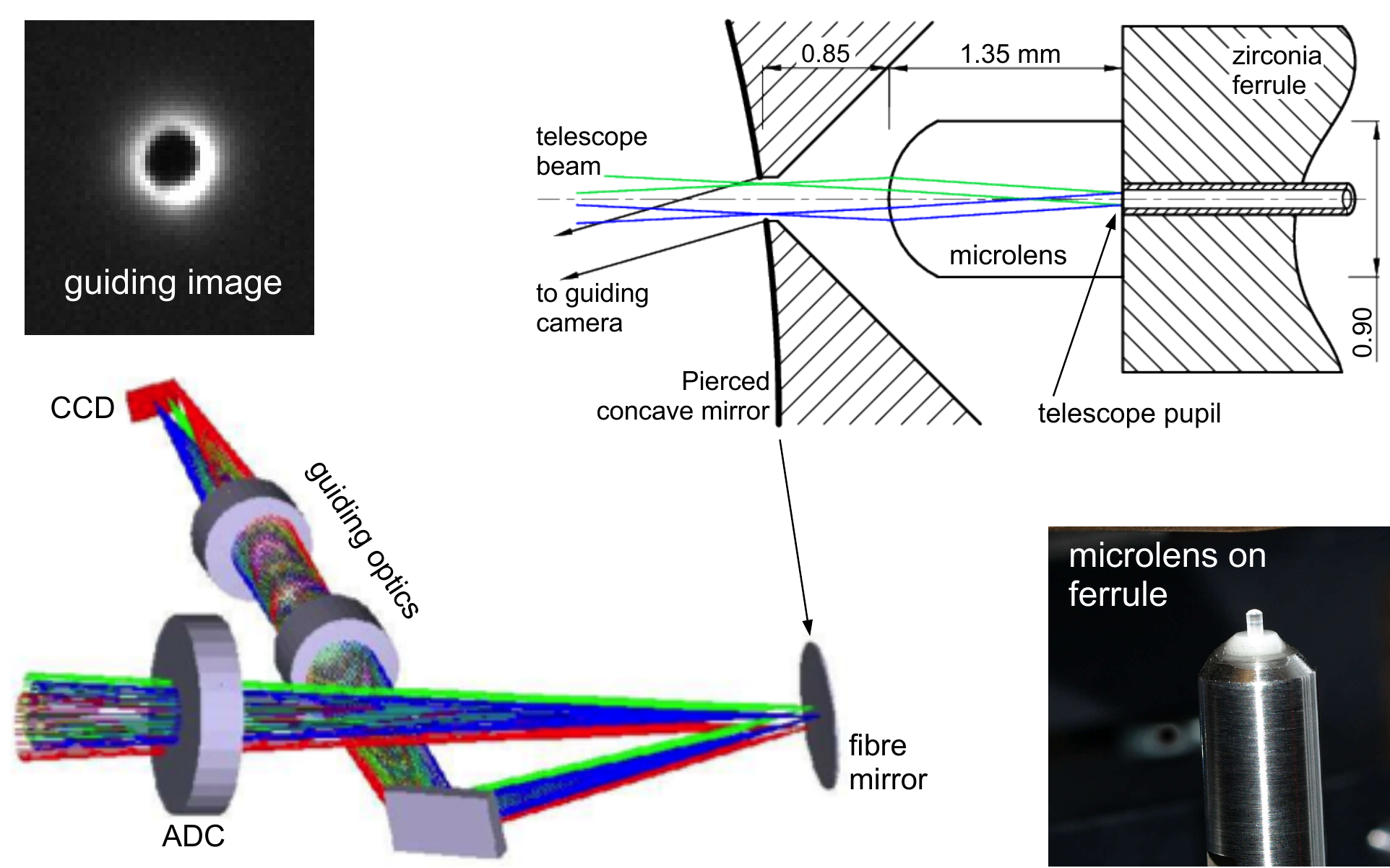}
\caption{\label{fig:fibreentrance} Top right: fibre entrance with pictures of a star centred on the fibre and the microlens cemented on the ferrule; bottom left: fibre viewer system (variable neutral density filter in front of guiding optics omitted).}
\end{figure}

HERMES does not have a fibre for sky subtraction. However, given the relatively small size of the Mercator telescope, this has never been perceived as an important shortcoming. Fig.\,\ref{fig:fibreentrance} shows the layout of the fibre entrance at the telescope side. A microlens, glued on top of a Zirconia ferrule, injects the the telescope pupil in the fibre. This fibre is mounted behind a small hole in a polished stainless steel mirror, located at the focal plane of the telescope. The  mirror is 8$^{\circ}$ inclined to reflect an image of the field to the guiding CCD. It has a concave surface to re-image the telescope pupil on the fibre viewer optics. Telescope guiding is performed on the light from the wings of the PSF that does not pass through the fibre hole in the mirror. To allow acquisition and guiding for a large range of brightness levels, a circular variable neutral density filter is used to attenuate from 0.1 to 100\%. 

The fibre entrance and fibre viewer are mounted in the telescope interface (Fig.\,\ref{fig:telescopeinterface}) that holds several other functions:  e.g. the calibration light projector to inject calibration light into  HRF or LRF, and the atmospheric dispersion corrector that provides a 4-step discrete correction of the differential atmospheric refraction. The length of the secondary spectrum remains well below 1 arcsec for zenithal angles up to 65$^{\circ}$.

\subsection{Fibre exit and exposure meter}
\begin{figure}
\includegraphics[angle=0, width=\linewidth]{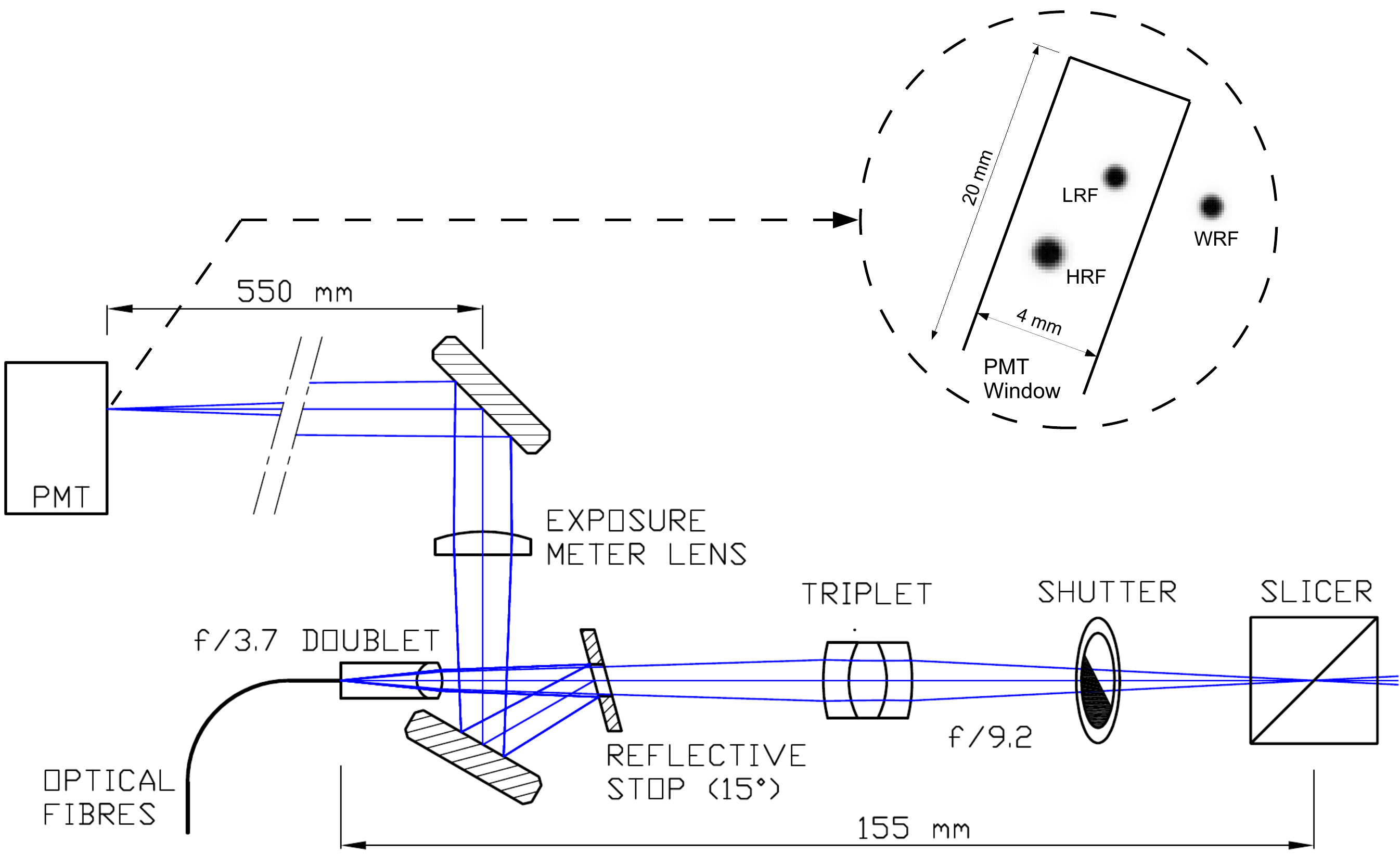}
\caption{\label{fig:fnoptics} Focal ratio adaptation optics with exposure meter.}
\end{figure}

At the spectrograph entrance, the focal ratio of the beam exiting the fibres is converted from $f$/3.7 to $f$/9.2 by a doublet and a symmetric triplet lens (Fig.\,\ref{fig:fnoptics}). The doublet is centred and cemented on both  HRF and LRF, as well as the wavelength reference fibre (WRF), in a common three-hole ferrule. An inclined and reflective aperture stop in between both lenses is exactly conjugated with the echelle grating. This stop intercepts the part of the beam that, due to focal ratio degradation (FRD), exits the fibres with a wider aperture than would be accepted by the grating. Two fold mirrors and a simple lens re-image the fibre exit surface on a photo-multiplier tube (PMT) that is used as flux monitor or exposure meter. In Fig.\,\ref{fig:fnoptics} is also shown how the image of the reference fibre is masked out to avoid contamination of the stellar flux with the Thorium lines.

In this configuration, the ratio between spectrograph and PMT flux is not absolutely constant but depends on FRD in the optical fibre.  When FRD increases, the PMT flux will increase too while the spectrograph flux will decrease. Hence, the accuracy of the exposure meter will be affected. To validate the suitability of this flux monitoring method, we measured the ratio between CCD and PMT flux while slewing the telescope 360$^{\circ}$ in azmiuth and 80$^{\circ}$ in elevation. A standard deviation of {\large$\sigma$}\,$\simeq$\,1\%  is found, but any correlation with telescope position is lacking. These measurements do not differ significantly from what we observe when the telescope is tracking slowly or is not moving at all, and FRD should be stable.

\subsection{Image slicer}
\begin{figure}
\includegraphics[angle=0, width=\linewidth]{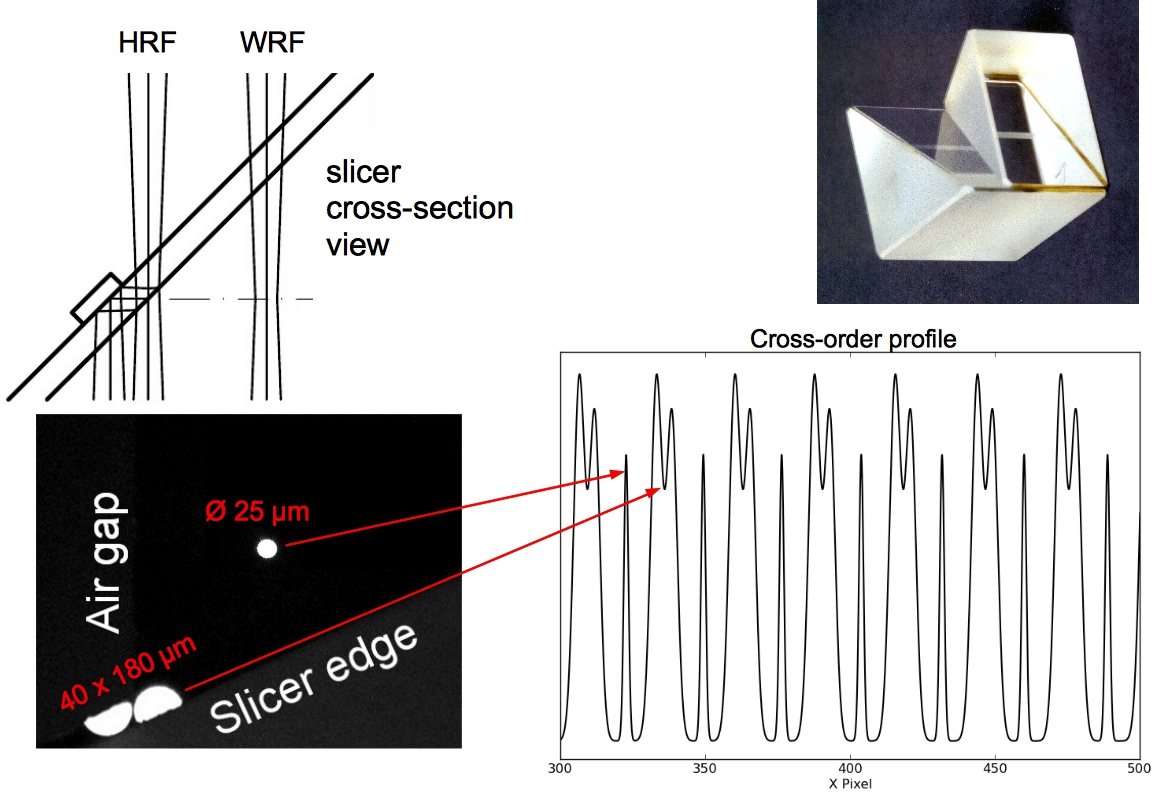}
\caption{\label{fig:slicer} Picture of image slicer (top right), view of slicer illuminated with 80-$\mu$m HRF and 25-$\mu$m WRF (left) an corresponding cross-order profile.}
\end{figure}

High spectral resolution (R\,=\,85\,000) is achieved by means of a two-slice image slicer that cuts the 80-$\mu$m HRF in two semi-circles (Fig.\,\ref{fig:slicer}). The slicer is of the FEROS type \citep{kaufer98b} and it is designed such that only HRF is sliced. The beams from LRF and WRF pass unaffected through the slicer. The sliced cross-order profile of HRF is three times wider than that of LRF and WRF. For this reason, it is currently not possible to interlace the HRF spectrum with a Thorium spectrum. 

However, as HRF offers both higher resolution and higher efficiency, this mode is used during 90\% of the HERMES observing time and observers are reluctant to sacrifice 30\% of both resolution and flux to obtain simultaneous Thorium exposures with LRF. For that reason, we plan an upgrade of the fibre link. The actual 60-$\mu$m reference fibre will be replaced by a new much smaller fibre with a diameter of only 25\,$\mu$m (Fig.\,\ref{fig:slicer}). The cross-order profile of this fibre is sufficiently narrow to fit the inter-order space of the sliced HRF up to at least 700\,nm. At longer wavelengths, the inter-order space becomes too small and therefore, the  Thorium-Argon flux redwards of 700\,nm will be removed by a short-pass filter to avoid contamination. At the same time, we plan to replace the LRF with an octagonal  80-$\mu$m sliced HRF. Several studies report improved scrambling and reduced FRD for octagonal fibres, compared to standard circular fibres \citep[e.g.][]{avila12, bouchy13}. With both types of fibre installed, we will be able to make an on-sky quantitative analysis and select the best option for future observations.

\subsection{Mechanics  and environment control}
\begin{figure}
\includegraphics[angle=90, width=\linewidth]{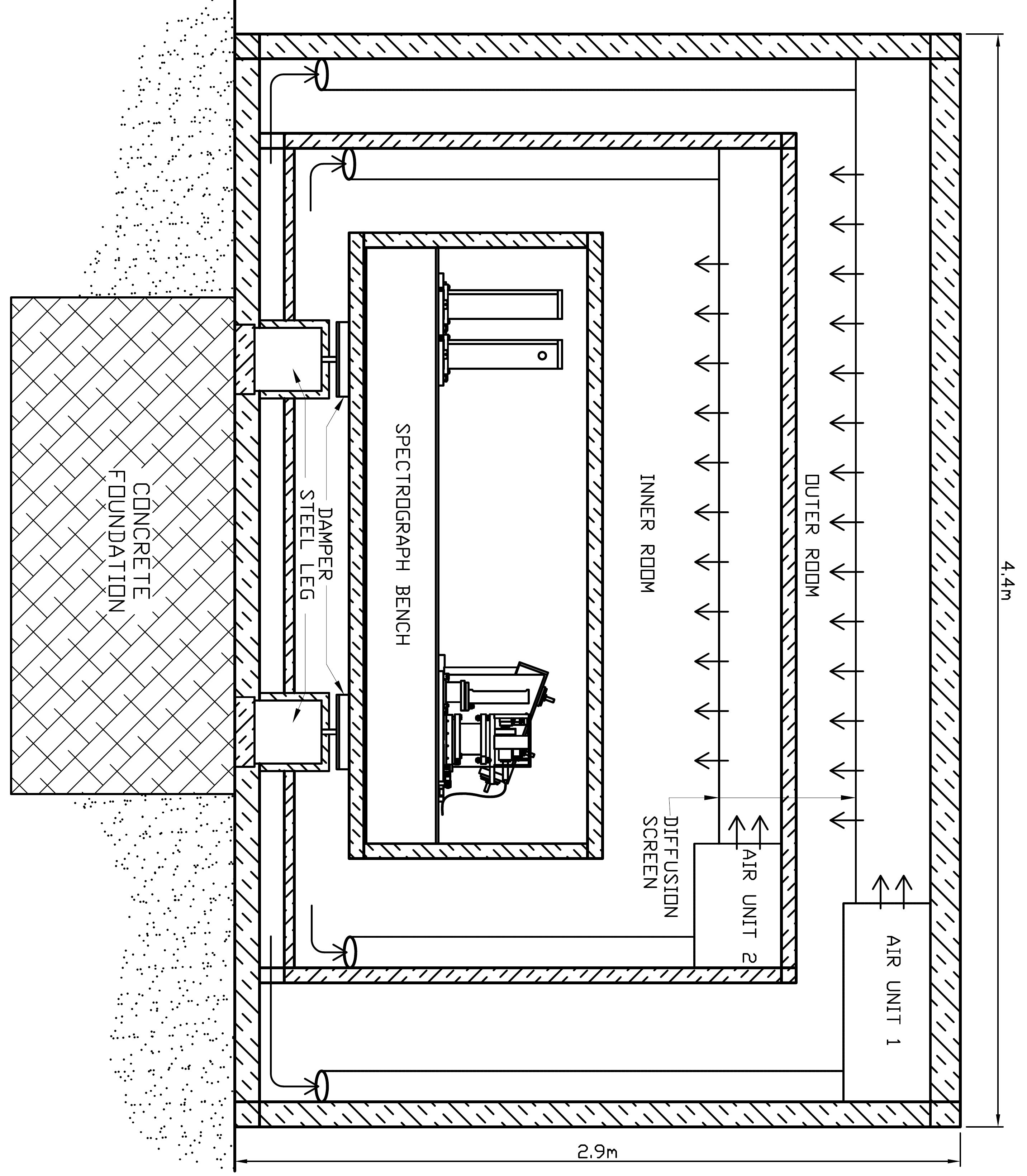}
\caption{\label{fig:room} Cross section of the HERMES room, the open arrows indicate the direction of  the forced air flow.}
\end{figure}

\begin{figure*}
\includegraphics[angle=0, width=\linewidth]{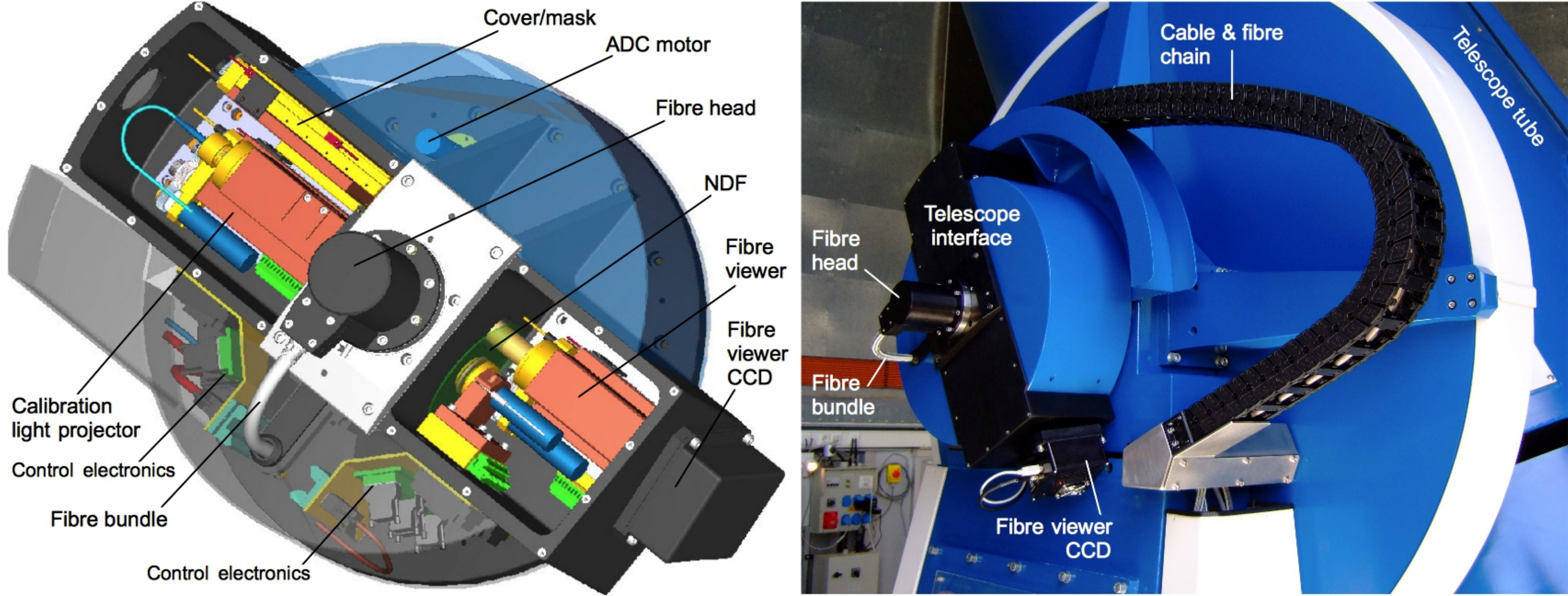}
\caption{\label{fig:telescopeinterface} Left: 3D-view of the telescope interface: right: telescope interface mounted at the Nasmyth focus of the Mercator telescope.}
\end{figure*}

The spectrograph is mounted on a standard optical bench (2400\,x\,1200\,x\,300\,mm) with passive vibration absorption. For improved vibration isolation, the bench is positioned on a solid concrete foundation, mechanically isolated from the rest of the telescope building (Fig.\,\ref{fig:room}). There are no moving parts on the optical bench that might compromise the mechanical stability. The only exception is a tiny shutter, installed at the fibre exit where the beam diameter is minimal (see Fig.\,\ref{fig:fnoptics}). Configuration of the spectrograph is completely done by the mechanisms in the telescope interface (Fig.\,\ref{fig:telescopeinterface}) and the calibration unit. The mounts of the optical components were all designed with long-term mechanical stability and thermal insensitivity in mind. 

\begin{figure}
\includegraphics[angle=0, width=\linewidth]{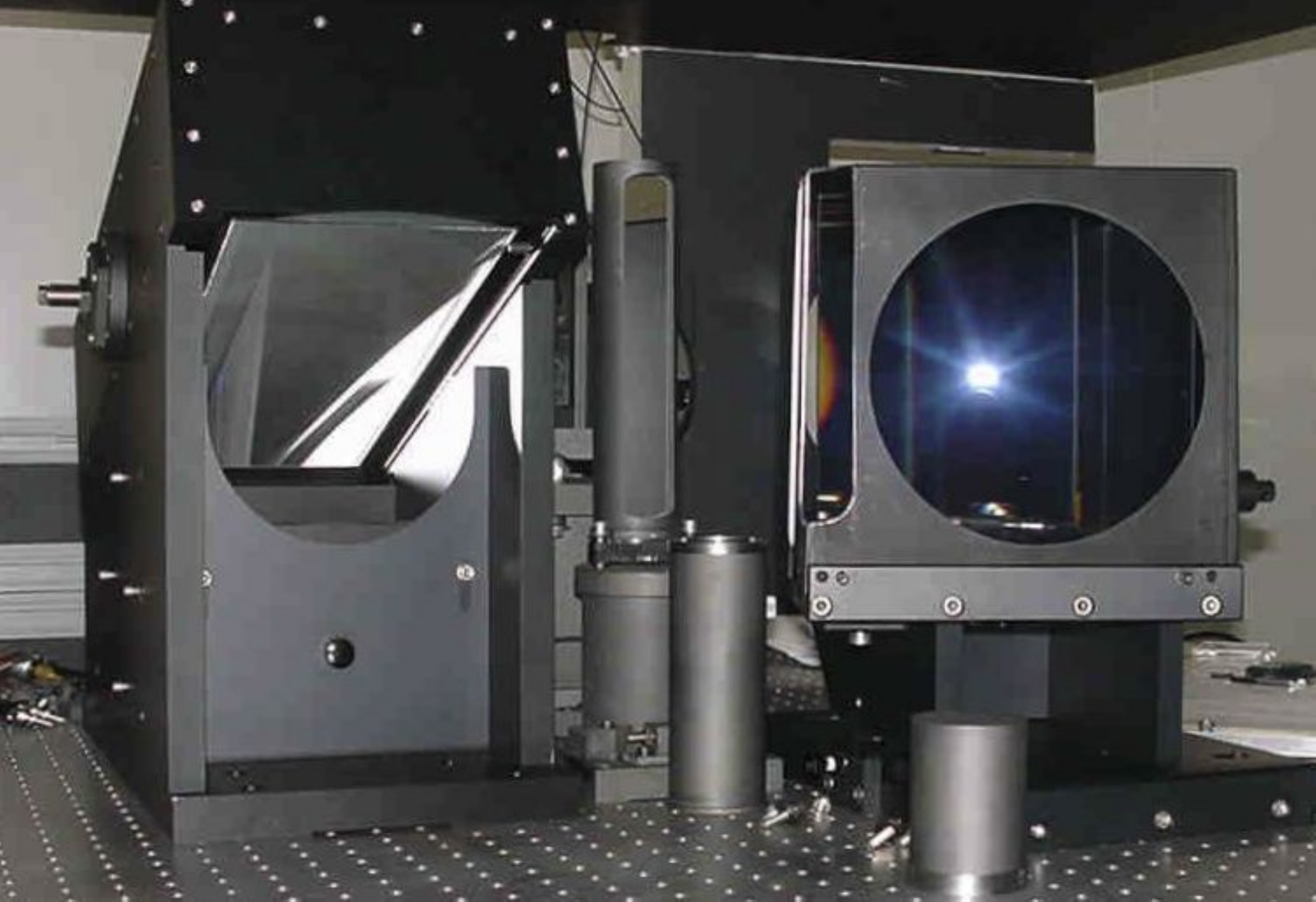}
\caption{\label{fig:photo-echelle} Picture of the echelle grating, fold mirror and prisms in their respective mounts on the spectrograph bench.}
\end{figure}

A stable thermal environment is essential for accurate radial velocity measurements. We achieve an extremely stable spectrograph temperature by means of a three-level isolation and a double thermal  control loop (Fig.\,\ref{fig:room}). The outer room is cooled to 13.5$^{\circ}$C with a stability of $\sigma$\,$<$\,0.1$^{\circ}$C. The inner room uses a 50\,W heater to raise the temperature to 18$^{\circ}$C. This leads to a temperature stability for the passive environment of the spectrograph bench of $\sigma$\,$<$\,0.01$^{\circ}$C (long term) or $\sigma$\,$<$\,0.001$^{\circ}$C (24 hours).

Atmospheric pressure is a second environmental parameter that affects the spectral stability. Pressure variations induce a shift at the detector of 0.05\,pixel\,hPa$^{-1}$ or a velocity  shift of 80\,ms$^{-1}$hPa$^{-1}$.  The walls of the outer spectrograph room have been reinforced to allow over-pressurization, with the idea of keeping the spectrograph's absolute pressure at a constant value.
However, we still have problems achieving a sufficiently hermetic room and the implementation of pressure control has been postponed. In retrospect, installing the spectrograph in a vacuum tank would have been a better choice, but this option was discarded  for budget reasons.

\section{Spectrograph performance overview}
\paragraph{Spectral resolution}
The nominal high resolution (HRF) of HERMES amounts to R\,=\,85\,000. At the center of the spectral orders, each high-resolution element is sampled by 2.3\,pixels. The anamorphosis of the spectrograph makes that at the blue end of the orders, this number decreases to 2\,pixels.  Because of the critical sampling, resolution drops to R\,=\,80\,000 in this region of the echellogram. These values compare very favourably to the geometric width of the sliced fibre that corresponds with only R\,=\,78\,000. 

In low-resolution mode (LRF), sampling is more relaxed (3\,--\,3.8\,pixels) along the full extent of the orders. As a result, spectral resolution has an almost constant value of R\,=\,63\,000. Again, the high quality of the HERMES optics is illustrated by comparing this number with the design value of R\,=\,52\,000, corresponding with the diameter of the fibre.

\begin{figure}
\includegraphics[angle=0, width=\linewidth]{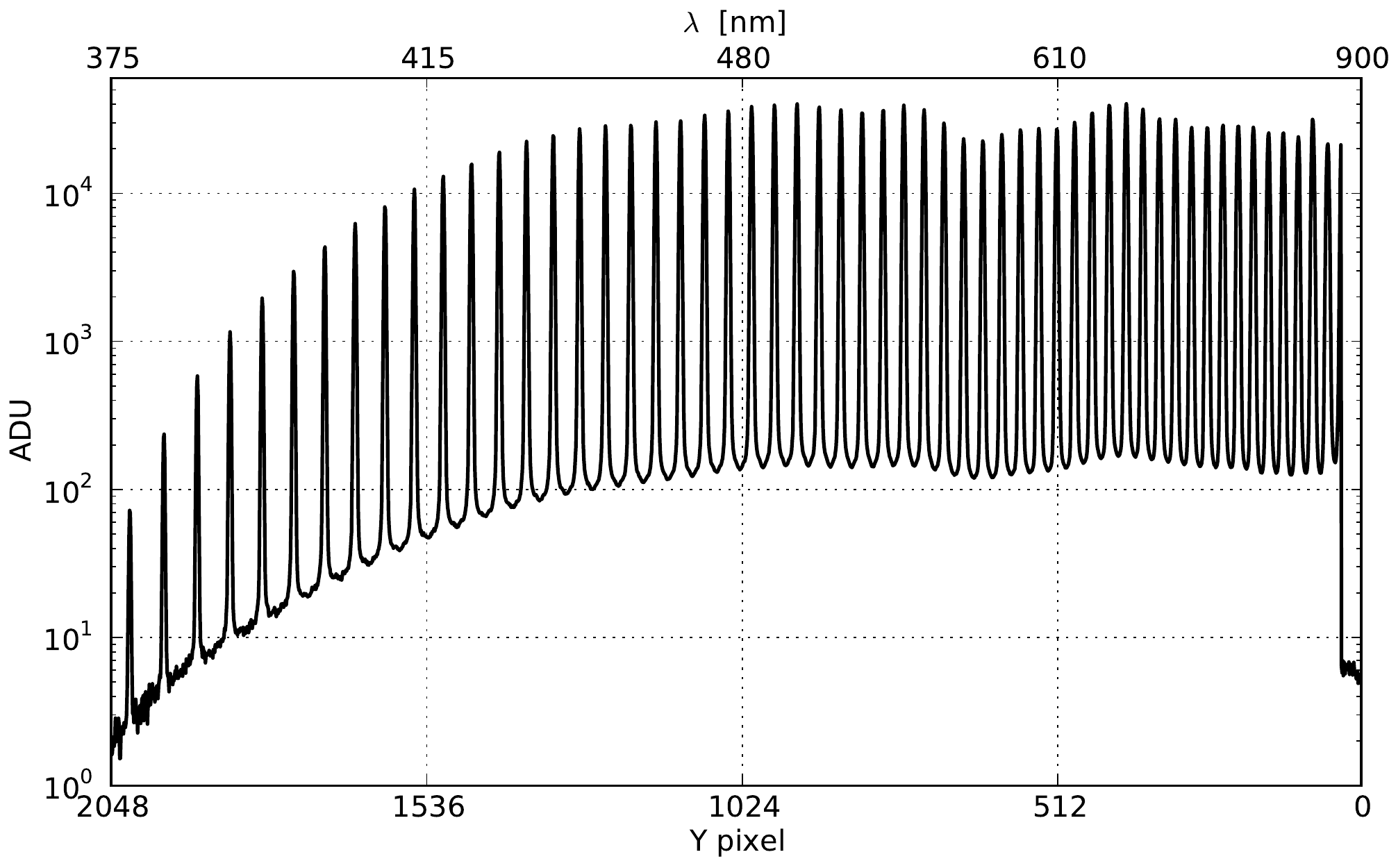}
\caption{\label{fig:straylight} Horizontal cross-cut through the centre of a flat-field spectrum, showing the inter-order stray light.}
\end{figure}

\paragraph{Stray light}
Thanks to very efficient baffling, allowed by the white-pupil layout, HERMES is a very \textit{clean} spectrograph. The spectra are virtually ghost-free and have only small stray light contamination. Fig.\,\ref{fig:straylight} shows a cross-order intensity tracing through the centre of a flat-field spectrum. The distribution of the scattered light is very local. The background in between the spectral orders hardly exceeds 100\,ADU while the peak flux in a large part of the spectrum amounts to more than 30\,000\,ADU. Except for the shortest wavelengths where the total signal is very small, the stray light signal in the inter-order pixels of both LRF and HRF spectra is approximately 0.1 per cent of the total flux in the adjacent orders.

\paragraph{Throughput}
High efficiency was a main driver throughout the HERMES design. As a result, throughput at blaze peak over the central part of the spectrum amounts to 28\% for the spectrograph alone in high-resolution mode (at other wavelength bands we find U:\,4\%, B:\,22\%, R:\,25\%, and 
I:\,17\%). If we include three mirror reflection of the telescope, total peak efficiency drops to 17.5\%. In LRF mode, these values have to be reduced by 30\% because of the throughput loss in the scrambler.

\begin{figure}
\includegraphics[angle=0, width=\linewidth]{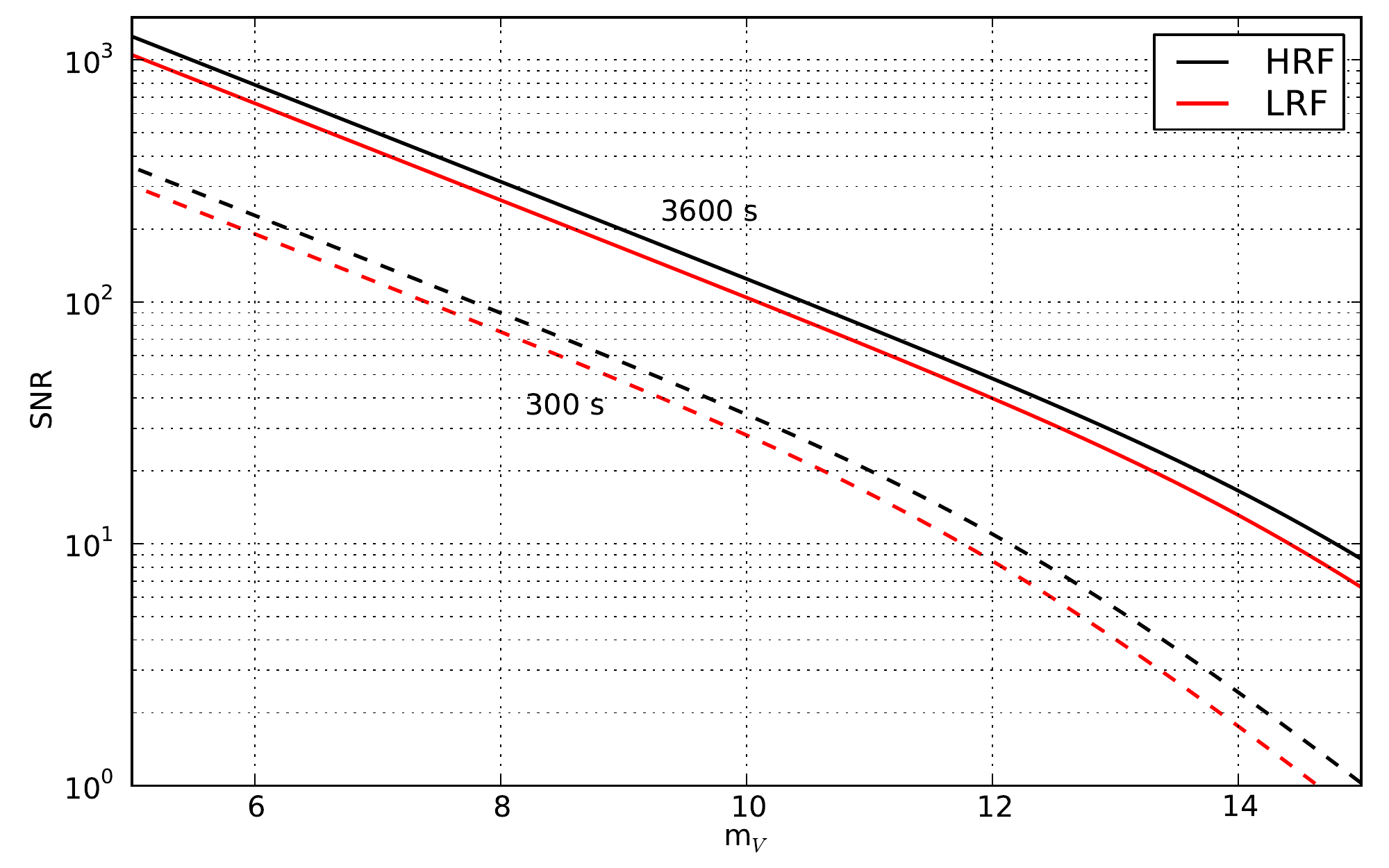}
\caption{\label{fig:snr} Estimated signal-to-noise ratio per pixel at 550\,nm and at blaze peak, as a function of stellar magnitude for a 5-minute (dashed) and a 1-hour (solid) exposure (seeing\,=\,1\,arcsec, F$_{\rm z}$\,=\,1).}
\end{figure}

With these throughput values and under good observing conditions (seeing\,=\,1\,arcsec, F$_{\rm z}$\,=\,1),  HERMES reaches in a one hour exposure a signal-to-noise ratio (SNR) of 100 from a m$_{\rm V}$\,=\,10.4 source  or a SNR of 10 with m$_{\rm V}$\,=\,14.7. These values are for HRF mode; in LRF mode the corresponding SNR values are 15\,--\,20\% lower. The estimated SNR is also shown graphically for a range of magnitudes in Fig.\,\ref{fig:snr}.

These limiting magnitudes and SNR values compare very favourable with those obtained from much larger telescopes. In high-resolution mode, HERMES delivers the same flux per \AA~as many other instruments on 2-m class telescopes. Thanks to the large sky aperture, HERMES can continue to operate with fairly high efficiency when seeing conditions degrade. Slit losses range from virtually zero at a seeing of 0.6\,arcsec to 5\% at 1.2\,arcsec and 25\% at 1.8\,arcsec. Under mediocre conditions, which unfortunately are not that rare at many observatories, high-resolution instruments on still larger telescopes can become very inefficient  because their sky apertures are small. Hence, in such circumstances, HERMES even stands its ground among these facilities

\begin{figure}
\includegraphics[angle=0, width=\linewidth]{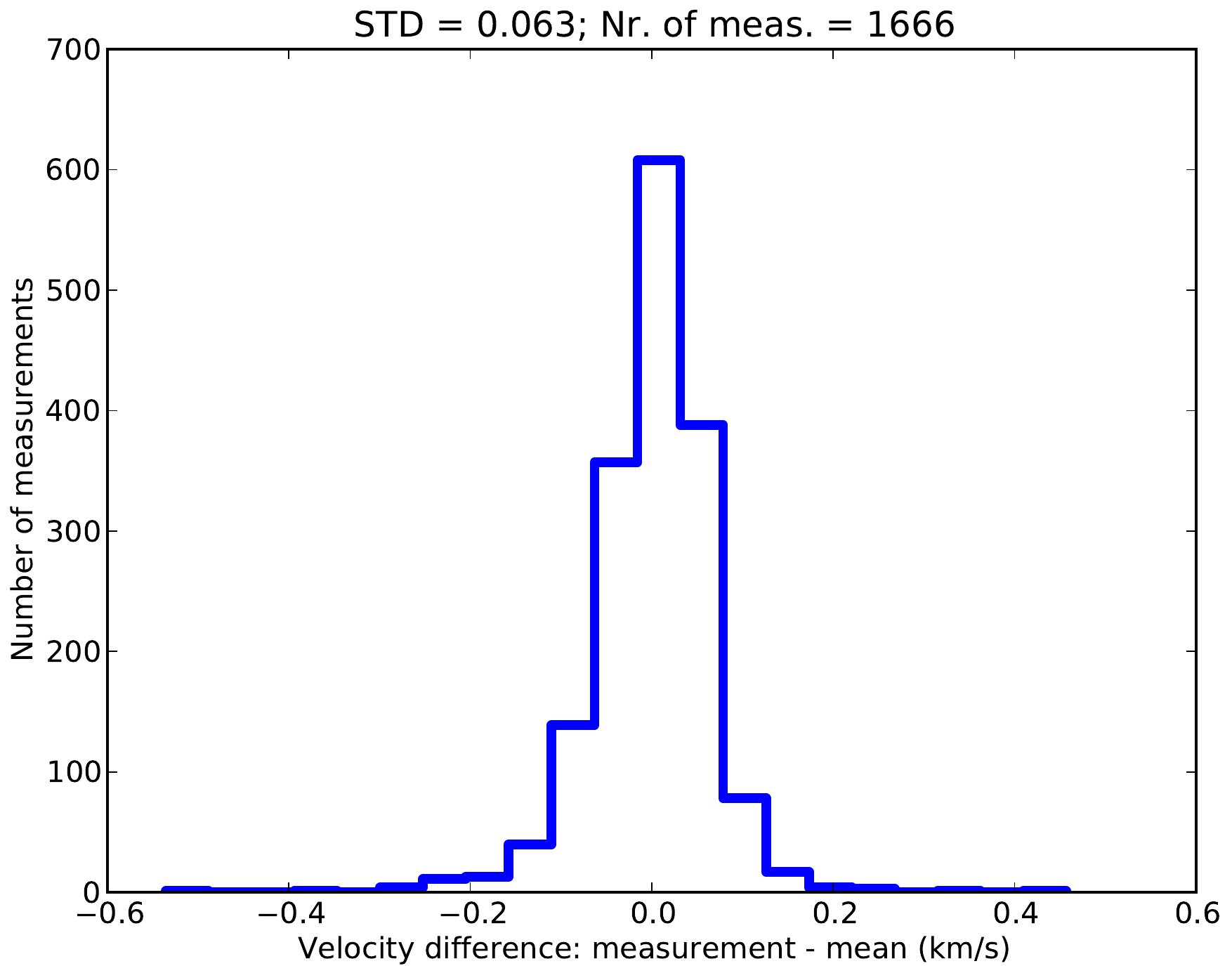}
\caption{\label{fig:RV_histo} Histogram with the distribution of 1666 radial velocity measurements of the IAU standard star HD\,164922 over four years of observations.}
\end{figure}

\paragraph{Stability and radial velocity accuracy}
During standard operation, we only acquire Thorium-Argon frames in the evening and morning of an observation night to calibrate the science frames in wavelength. Nightly we also obtain spectra of a few IAU velocity standards. Due to the variable environmental conditions, we find a standard deviation of 63\,m/s on the radial velocity measurements of 1666 exposures of the IAU standard HD\,164922, taken over a period of more than four years (Fig.\,\ref{fig:RV_histo}). 

The short-term stability of HERMES is strongly affected by the atmospheric pressure variations during the night. However, when the the short-term instrumental drift is corrected by the simultaneous Thorium-Argon exposure in LRF mode or by regularly interleaving the science exposures with a Thorium-Argon reference spectrum (e.g. 0.5 hour intervals) in the HRF mode, we  find a standard deviation on the velocity of 2\,\ms\ for LRF or 2.5\,\ms\ for HRF mode \citep{Raskin11}.

\begin{figure*}
\includegraphics[angle=0, width=\linewidth]{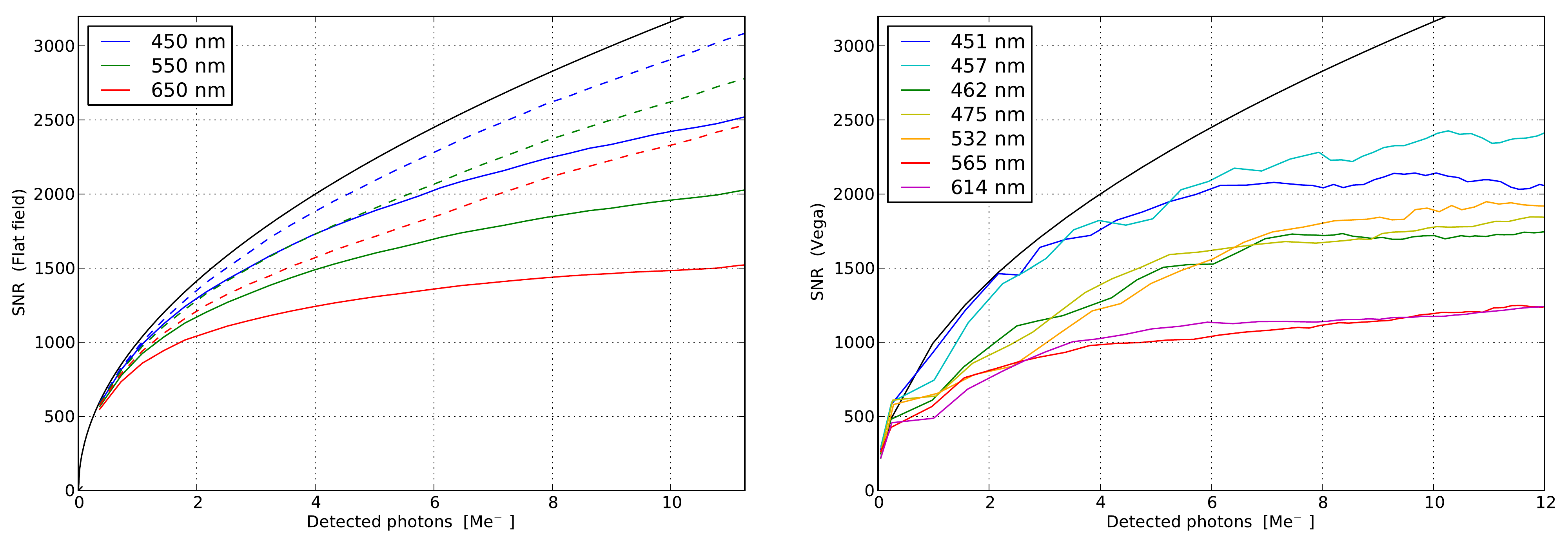}
\caption{\label{fig:modalnoise} Effect of modal noise on SNR, solid black curve shows theoretical SNR. \textit{Left}: measured  SNR of summed flat-field spectra with flat-fielding done without (dashed) or after (solid) disturbing the fibres by moving the telescope; \textit{right}: measured SNR of summed spectra of Vega at various wavelengths (colours indicated in legend).  }
\end{figure*}

\paragraph{Modal fibre noise and SNR limitations}
Several studies \citep[e.g.][]{Baudrand01, Grupp03, lemke11} report about the SNR limitations of fibre-fed spectrographs imposed by modal fibre noise. As some HERMES science programs target very high SNR of bright targets, the effect of modal noise was also investigated for HERMES. First we measured the SNR of 32 summed flat field (FF) spectra, divided by another series of 32 FF spectra, taken under identical conditions. Then we moved the telescope to a completely different position, took a new series of 32 FF spectra and divided them as well by the original FF spectra. The results are shown in the left panel of Fig.\,\ref{fig:modalnoise}. The resulting SNR is clearly smaller than the theoretical SNR, purely set by photon noise statistics. As expected from modal noise, the SNR drop increases with wavelength. When in between two series of FF spectra,  the fibres are disturbed by moving the telescope, modal noise is  strongly affected. This results in a further decrease of the SNR.

The right panel of Fig.\,\ref{fig:modalnoise} shows the SNR measured in the continuum part of the sum of 40 Vega spectra. We find that the SNR that practically can be achieved with HERMES is limited to $\sim$1000 in the red part of the spectrum, increasing to 2000\,--\,2500 at blue wavelengths. 

Finally, we also verified that modal fibre noise is only very marginally improved by the use of the optical scrambler in LRF mode. As was expected, double optical scrambling does not solve the SNR limitations imposed by modal noise. Installing a mechanical scrambler or fibre shaker might provide a solution to the modal noise SNR limitations.

\section{Results}
The strongest scientific asset of the Mercator+HERMES combination lies
in the ability to provide stellar astrophysicists with high-quality
time series of high-resolution spectra and this with a wide range of
sampling times.  It is one of the rare telescope--instrument
combinations which can provide also very long monitoring programmes,
ideal for e.g.  asteroseismology and binary star research. Moreover,
the high-resolution spectra provide an ideal complement to
space missions like CoRot and \textit{Kepler}, which only provide photometric
time series of white light or light obtained with broad spectral
filters.  

Since the start of science observations in April 2009, HERMES has collected more than 42\,000 science spectra during a total of 1250 observing nights. Till now, this resulted in close to 100 scientific papers exploiting
this facility, going from the discovery of long-period eccentric
binaries among sub-dwarf B stars \citep{vos2012,oestensen2012}, the
discovery of the binary properties of red giant pulsators in the
\textit{Kepler} field and also often in eccentric systems \citep{frandsen2013}, the mode identification of single and binary pulsators
\citep[e.g.][]{degroote2012,debosscher2013,tkachenko2013} and the circumstellar gas flows in binary stars with interacting winds or accretion flows \citep{lobel2013, gorlova2012}. These are only some illustrative example of science programmes
which require the unique monitoring capabilities of HERMES on the Mercator telescope.

\section{Conclusions}
The examples of the previous section illustrate that today, with  extremely large telescopes dawning at the horizon, a small but dedicated telescope and equipped with an efficient high-resolution spectrograph, is still capable of producing valuable and even unique science. Moreover, this can be done at just a fraction  of the cost of large facilities.
HERMES on the Mercator telescope is an ideal combination to achieve this: a very efficient instrument featuring high resolution (R\,=\,85\,000), large single-shot coverage (380\,--\,900\,nm) and excellent stability, matched with a flexible 1.2-m telescope with almost permanent and long-term availability. 
Accordingly,  the productivity of both HERMES and the Mercator telescope forms a key element in many of the science programs that are conducted at the Institute of Astronomy of the KU\,Leuven or by our consortium partners. To further improve HERMES' performance, an upgrade of the optical fibre link is being planned.

Mercator is not an open telescope. This means that HERMES is only available to the community through collaboration with one of the consortium members or through the Spanish CAT time allocation committee that allocates 20\% of the observing nights of the Roque de los Muchachos observatory.

\vspace{1mm}

\acknowledgements
The HERMES project and team acknowledge support from the Fund for Scientific Research of
Flanders (FWO) under grants G.0472.04, G.0C31.13 and G0.470.07, from
the Research Council of K.U.Leuven under grant GST-B4443,
from the Fonds National de la Recherche Scientifique under contracts IISN\,4.4506.05 and FRFC\,2.4533.09,
and financial support from Lotto (2004) assigned to the Royal Observatory of 
Belgium to contribute to the hardware of the spectrograph.

The authors thank the HERMES and Mercator teams for their dedication and essential contributions to the success of the HERMES project.


\end{document}